\documentstyle[amssymb,psfig]{mn2e}

\title[Star cluster ``infant mortality'' in the SMC]{Star cluster
``infant mortality'' in the Small Magellanic Cloud (Redivivus)}

\author[R. de Grijs and S. P. Goodwin]{Richard
de Grijs$^{1,2}$\thanks{E-mail: R.deGrijs@sheffield.ac.uk} and Simon
P. Goodwin$^1$\\ 
$^1$ Department of Physics \& Astronomy, The University of Sheffield,
Hicks Building, Hounsfield Road, Sheffield S3 7RH\\ 
$^2$ National Astronomical Observatories, Chinese Academy of
Sciences, 20A Datun Road, Chaoyang District, Beijing 100012,
China}

\date{Received date; accepted date}
\pubyear{2007}

\begin{document}
\maketitle

\begin{abstract}
The early evolution of star clusters in the Small Magellanic Cloud
(SMC) has been the subject of significant recent controversy,
particularly regarding the importance and length of the earliest,
largely mass-independent disruption phase (referred to as ``infant
mortality''). Here, we take a fresh approach to the problem, using an
independent, homogeneous data set of $UBVR$ imaging observations, from
which we obtain the SMC's cluster age and mass distributions in a
self-consistent manner. We conclude that the (optically selected) SMC
star cluster population has undergone at most $\sim 30$ per cent
($1\sigma$) infant mortality between the age range from about $(3-10)$
Myr, to that of approximately $(40-160)$ Myr. We rule out a 90 per
cent cluster mortality rate per decade of age (for the full age range
up to $10^9$ yr) at a $> 6 \sigma$ level. We independently affirm this
scenario based on the age distribution of the SMC cluster sample.
\end{abstract}


\section{Introduction}
\label{intro.sec}

The early evolution of the star cluster population in the Small
Magellanic Cloud (SMC) has been the subject of considerable recent
attention and vigorous debate (e.g., Rafelski \& Zaritsky 2005;
Chandar, Fall \& Whitmore 2006; Chiosi et al. 2006; Gieles, Lamers \&
Portegies Zwart 2007). The key issue of contention is whether the
SMC's star cluster system has been subject to the significant early
cluster disruption processes observed in ``normal'', interacting and
starburst galaxies commonly referred to as ``infant mortality'' and
``infant weight loss''. Chandar et al. (2006) argue that the SMC has
been losing up to 90 per cent of its star clusters per decade of age,
at least for ages from $\sim 10^7$ up to $\sim 10^9$ yr, whereas
Gieles et al. (2007) conclude that there is no such evidence for a
rapid decline in the cluster population, and that the decreasing
number of clusters with increasing age is simply caused by fading of
their stellar populations. They contend that the difference between
their results was due to Chandar et al. (2006) assuming that they were
dealing with a mass-limited sample, whereas it is actually
magnitude-limited. In fact, this is not entirely correct; Chandar et
al. (2006) analyse the full magnitude-limited sample and conclude that
it is approximately surface-brightness limited. They then compare the
cluster age distribution of the full sample (expressed in units of
${\rm d}N_{\rm cl} / {\rm d}t$, i.e., the number of clusters per unit
time period) to that of a subsample for masses $\ge 10^3$ M$_\odot$
(which they do not analyse in the same manner), and suggest both to be
similar, although the latter is much flatter\footnote{Although Chandar
et al. (2006) suggest that their sample is roughly mass limited, they
also note that the mass-limited subsample, constrained to clusters
with masses $\log( M_{\rm cl} / {\rm M}_\odot ) \ge 3.5$, shows a
flatter age distribution.}, hence giving rise to the discrepancy
between their results and those of Gieles et al. (2007). Both studies
are based on the same data set, the Magellanic Clouds Photometric
Survey (MCPS; Zaritsky, Harris \& Thompson 1997).

The main contribution of this paper to the ongoing debate is two-fold:
(i) We revisit the SMC's early star cluster evolution using an
alternative approach; and (ii) we use independently obtained ages and
masses based on an independent data set, i.e., the $UBVR$ photometric
survey of the Magellanic Clouds by Massey (2002), originally analysed
by Hunter et al. (2003). We conclude that there is indeed only
marginal evidence for infant mortality in the SMC star cluster sample,
supporting the careful analysis of Gieles et al. (2007). In Sect.
\ref{infant.sec} we first briefly introduce the concept of cluster
infant mortality. We discuss our observational data and the basic
analysis leading to the age and mass estimates in Sect.
\ref{data.sec}. In Sect. \ref{analysis.sec} we justify our choice of
age ranges to construct cluster mass functions (CMFs). Finally, in
Sect. \ref{discussion.sec}, we present our case for the absence of
significant cluster infant mortality.

\section{Cluster infant mortality}
\label{infant.sec}

Observations of increasing numbers of interacting and starburst
galaxies, including the Antennae system, M51 and NGC 3310, show a
significantly larger number of young ($\lesssim 10-30$ Myr) star
clusters than expected from a simple extrapolation of the cluster
numbers at older ages, taking into account the observational
completeness limits and the effects of sample binning, and under the
additional, simplifying assumption that the star cluster formation
rate (CFR) has been roughly constant over the host galaxy's history
(e.g., de Grijs et al. 2003b; Whitmore 2004; Bastian et al. 2005;
Fall, Chandar \& Whitmore 2005; Mengel et al. 2005; Chandar et
al. 2006; see also Whitmore, Chandar \& Fall 2007 for a presentation
of earlier results, and de Grijs \& Parmentier 2007 for a review).
This significant overdensity remains, even in view of the presence of
a recent burst of star cluster formation in many of these galaxies.

These observations have prompted a flurry of activity in the area of
star cluster disruption processes. This has led to suggestions that
cluster systems appear to be affected by a disruption mechanism that
acts on very short time-scales ($\lesssim 10-30$ Myr) and which may be
mass-independent -- at least for masses in excess of $\sim 10^4$
M$_\odot$ (e.g., Fall et al. 2005; Bastian et al. 2005; Fall
2006). This fast disruption mechanism, which is thought to effectively
remove around 50 (Goodwin \& Bastian 2006; although their sample is
very probably biased), 70 (Bastian et al. 2005; Mengel et al. 2005) or
even 90 per cent (Lada \& Lada 1991; Whitmore 2004; Whitmore et
al. 2007) of the youngest clusters from a given cluster population, is
thought to be the rapid removal of the intracluster gas on a
time-scale of $\sim 5$ Myr, the signatures of which have been seen in
several clusters (Bastian \& Goodwin 2006). The observational effect
resulting from this rapid gas removal has been coined cluster ``infant
mortality'' (Lada \& Lada 2003); it was originally reported in the
context of the number of very young embedded clusters, compared to
their older, largely gas-free counterparts in the Milky Way.

The general consensus emerging from recent studies is that rapid gas
removal from young star clusters is likely to leave the clusters
super-virial and hence lead to the rapid disruption of many clusters
(see, e.g., Goodwin 1997a,b; Boily \& Kroupa 2003a,b; Goodwin \&
Bastian 2006; see also de Grijs \& Parmentier 2007 for a review). This
leaves surviving clusters more susceptible to destruction (Vesperini
\& Zepf 2003; Bastian et al. 2005; Fall et al. 2005).

As described by Goodwin \& Bastian (2006, and references therein) the
effect of gas removal is to rapidly decrease the potential well in
which the stars reside. The cluster will expand in an attempt to
return to virial equilibrium. If the virial ratio of the stars {\em
after} gas expulsion is $\ga 3$ the cluster will be unable to return
to an equilibirum and will be destroyed\footnote{This is usually given
in terms of a star-formation efficiency of $\sim 30$ per
cent. However, as noted by Goodwin \& Bastian (2006), it is the virial
ratio of the stars after gas expulsion that is the crucial parameter,
and this can only be related to the star-formation efficiency in a
simple way if the stars and gas were initially in virial equilibrium
with one another. For this reason, Goodwin \& Bastian (2006) use the
term ``effective star formation efficiency''.}. Clusters with a higher
effective star-formation efficiency than around $30$ per cent will
survive, but may undergo significant ``infant weightloss'' (loosing in
excess of 50 per cent of their initial mass in some cases). The
signature of infant weightloss has been observed in several young
clusters (Bastian \& Goodwin 2006). The time-scale over which a
cluster will be destroyed, or attain a new (lower-mass) equilibrium
configuration is $10 - 40$ Myr (depending on the effective
star-formation efficiency and the cluster mass).

The $10 - 40$ Myr time-scale of gas removal-induced infant mortality
and infant weightloss has important consequences for the analysis and
interpretation of the data in this paper. Clusters undergoing
expansion will have decreasing surface brightnesses, thus reducing
their chances of being detected as they grow older. However, some
clusters will recollapse after $10 - 40$ Myr, which may bring them
back into the sample. In addition, the speed at which clusters are
lost from the sample would be expected to depend on their (initial)
mass. Lower-mass clusters which are initially only just above the
detection limit will drop out of the sample very quickly, whilst
larger clusters may remain in the sample (albeit with a lower surface
brightness) for longer. It is also almost impossible (without
extensive observations of the surface brightness profiles and/or
dynamical state of the cluster) to determine which clusters that are
present in this age range will survive gas expulsion, and which are
headed for destruction. Thus, the interpretation of the numbers and
mass function of clusters in the age range $10 - 40$ Myr is fraught
with problems (in addition to these ``physical'' problems, age
determinations for clusters in this age range also cause problems; see
below).

\section{A homogeneous photometric database}
\label{data.sec}

\begin{figure}
\psfig{figure=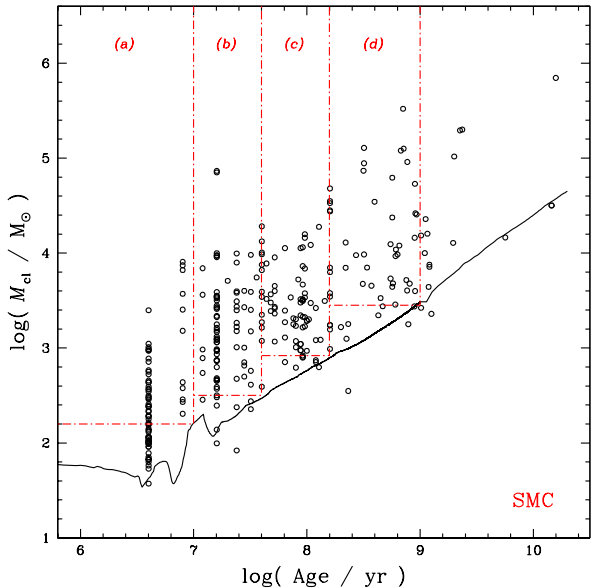,width=\columnwidth}
\caption{\label{smcamd2.fig}Distribution of the SMC clusters in the
log(age) vs. log(mass) plane. Overplotted is the expected detection
limit based on stellar population synthesis for a 50 per cent
completeness limit of $M_V = -4.5$ mag, assuming no extinction. For a
nominal extinction of $A_V = 0.08$ mag (assuming the Calzetti
attenuation law), the detection limit will shift to higher masses by
$\Delta \log( M_{\rm cl}/ M_\odot ) = 0.03$, which is well within the
uncertainties associated with our mass determinations (see de Grijs \&
Anders 2006). The features around 10 Myr are caused by the appearance
of red supergiants in the models. The age limits used to generate the
different panels in Fig. \ref{smcclfs4.fig} are shown as the vertical
dash-dotted lines; the various subsets are also cross-linked between
the figures using the panel indications from Fig. \ref{smcclfs4.fig}.
The horizontal dash-dotted lines indicate the 50 per cent completeness
limits in mass for each of the age-selected subsamples.}
\end{figure}

The basis for our detailed re-analysis of the SMC star cluster system
is provided by the $UBVR$ broad-band spectral energy distributions
(SEDs) of Hunter et al. (2003), based on Massey's (2002) CCD survey of
the Magellanic Clouds.

In a series of recent papers, we developed a sophisticated tool for
star cluster analysis based on broad-band SEDs, {\sc AnalySED}, which
we tested extensively both internally (de Grijs et al. 2003a,b; Anders
et al. 2004) and externally (de Grijs et al. 2005), using both
theoretical and observed young to intermediate-age ($\lesssim 3 \times
10^9$ yr) star cluster SEDs, and the {\sc galev} ``simple'' stellar
population (SSP) models (Kurth et al. 1999; Schulz et al. 2002). The
accuracy has been further increased for younger ages by the inclusion
of an extensive set of nebular emission lines, as well as gaseous
continuum emission (Anders \& Fritze-v. Alvensleben 2003). We
concluded that the {\it relative} ages and masses within a given
cluster system can be determined to a very high accuracy, depending on
the specific combination of passbands used (Anders et al. 2004). Even
when comparing the results of different groups using the same data
set, we can retrieve any prominent features in the cluster age and
mass distributions to within $\Delta \langle \log( {\rm Age / yr} )
\rangle \le 0.35$ and $\Delta \langle \log( M_{\rm cl} / {\rm M}_\odot
) \rangle \le 0.14$, respectively (de Grijs et al. 2005), which
confirms that we understand the uncertainties associated with the use
of our {\sc AnalySED} tool to a very high degree.

In de Grijs \& Anders (2006) we presented newly and homogeneously
redetermined age and mass estimates for the entire Large Magellanic
Cloud (LMC) star cluster sample covered by the Massey (2002) data.
Based on the comparison of our results in de Grijs \& Anders (2006)
with those published previously in a range of independent studies
(mostly based on spectroscopic or isochrone analyses), and
additionally on a detailed assessment of the age-metallicity and
age-extinction degeneracies, we concluded that our broad-band SED fits
yield reliable ages, with statistical {\it absolute} uncertainties
within $\Delta\log( \mbox{Age/yr}) \simeq 0.4$ overall. Here, we
extend this to the SMC cluster sample, using the same age-dating
technique as described above.

Our cluster age and mass determinations assume an average metallicity
of $Z = 0.008$ (where Z$_\odot = 0.020$), and a mean foreground
extinction $E(B-V) = 0.08$ mag. We will justify both of these choices
below. In Fig. \ref{smcamd2.fig} we show the distribution of our SMC
cluster sample in the log(age) vs. log(mass) plane; the adopted 50 per
cent completeness limit is overplotted. We have also indicated the
regions in this plane from which we have drawn statistically complete
subsamples, which we will discuss in detail in Sect.
\ref{analysis.sec}.

The determination of the 50 per cent completeness limit of the SMC
cluster data is in essence based on a close inspection of the cluster
photometry contained in Hunter et al.'s (2003) fig. 11. These authors
selected their sample from the catalogue of Pietrzy\'nski et
al. (1998), matched to the observational field of view of the Massey
(2002) data. Therefore, our completeness is that of this catalogue;
Hunter and her team did not quantify the completeness levels
themselves (D. Hunter, priv. comm.), although they discuss an observed
fading limit. However, for our analysis it is important to understand
the sample incompleteness affecting our observations. As such, we
adopted the conservative approach that the present-day SMC cluster
luminosity function (CLF; see Hunter et al.'s fig. 11) is best
represented by a power-law function in luminosity. Based on this
assumption, we used the same observational data as used by Hunter et
al. (2003) to determine the 50 per cent completeness limit at $M_V
\sim -4.5 \pm 0.2$ mag (based on a power-law fit to the clusters
brighter than $M_V = -5$ mag; varying this lower limit by a few tenths
of a magnitude does not result in a significant change), i.e., at the
same level as Hunter et al.'s (2003) observed fading limit. We note
that if the underlying CLF is {\it not} a single power law over the
entire observed luminosity range, the limit we adopt following this
approach is in fact a lower limit. In the latter case the observations
will likely be more complete than estimated here. Since the adopted 50
per cent completeness limit describes the lower envelope of the
distribution of our SMC cluster sample very well, we are confident
that our approach is reasonable. In addition, we point out that a
variation in the magnitude limit of 0.2--0.4 mag will shift the mass
limit by at most 0.1--0.2 dex, which clearly is still within our range
of uncertainties. The magnitude of the shift expected when going from
the 50 to the 90 per cent completeness limit in the SMC disc is of the
same order, $\sim 0.5$ mag. Finally, we point out that it is most
likely that the completeness of our SMC cluster sample is in fact
determined by the $U$-band observations. From a direct comparison of
the $U$-band and the $V$-band data, we derive a 50 per cent
completeness in the $U$ band at $M_U = -5.0 \pm 0.3$ mag.

Chiosi et al. (2006) recently analysed the star-formation history in
the SMC in detail using an independently selected star cluster
sample. Where possible, they derive the extinction towards individual
clusters based on colour-magnitude diagram analysis, and otherwise
assume a mean extinction $E(B-V) = 0.08$ mag, following Tumlinson et
al. (2002) and Hunter et al. (2003; see also Rafelski \& Zaritsky
2005). We have adopted the same average extinction value to our SMC
cluster sample, using the Calzetti attenuation law (Calzetti 1997,
2001; Calzetti et al. 2000; Leitherer et al. 2002) with $R_V =
4.05$.

Rafelski \& Zaritsky (2005) obtained SMC cluster ages of a small
cluster sample on the basis of three sets of models, for metallicities
of $Z=0.001, 0.004$ and 0.008. They concluded that some of the
lowest-metallicity models could be rejected and adopted $Z=0.008$ as
an appropriate mean metallicity for their SMC cluster sample. Chiosi
et al. (2006) also adopted this metallicity, but for their younger
sample clusters. For the older ($\gtrsim 1-2$ Gyr) clusters, they
assumed $Z=0.004$, as was also done by Hunter et al. (2003). However,
as shown in Fig. \ref{smcamd2.fig}, the large majority of our sample
clusters (and in particular the subpopulations we will analyse in more
detail below) are younger than $\sim 1$ Gyr, so that we adopt
$Z=0.008$ as the mean metallicity for our SMC cluster sample.

\section{The cluster mass function}
\label{analysis.sec}

In de Grijs \& Anders (2006) we presented the cumulative CMFs of the
LMC star clusters younger than certain age limits. We concluded that
while the older cluster (sub)samples are characterised by CMF slopes
consistent with the $\alpha \simeq 2$ slopes generally observed in
young star cluster systems -- where $\alpha$ is defined as $N(M_{\rm
cl}) \propto M_{\rm cl}^{-\alpha}$ -- the youngest mass and
luminosity-limited LMC cluster subsets show shallower slopes (at least
below masses of a few $\times 10^3$ M$_\odot$). We noted that we could
not disentangle the unbound from the bound clusters at the youngest
ages. This is what we set out to do here for the SMC cluster system.

\begin{figure*}
\hspace{0.75cm}
\psfig{figure=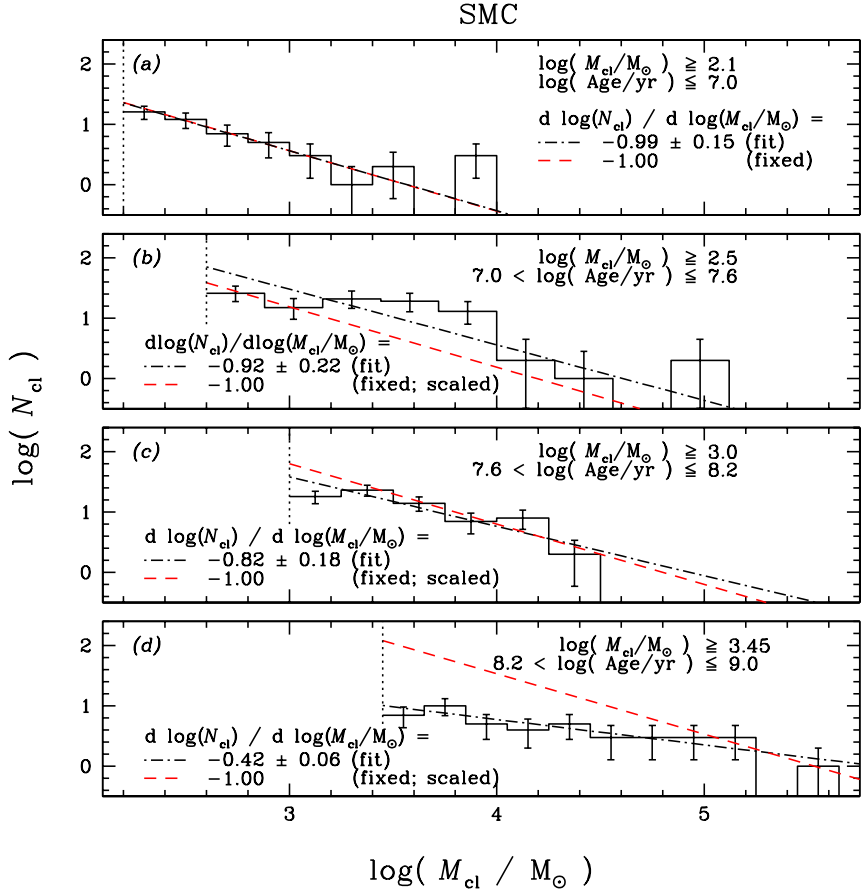,width=12cm}
\caption{\label{smcclfs4.fig}CMFs for statistically complete SMC
cluster subsamples. Age and mass ranges are indicated in the panel
legends; the vertical dotted lines indicate the lower mass (50 per
cent completeness) limits adopted. Error bars represent simple
Poissonian errors, while the dashed lines represent CMFs of slope
$\alpha = 2$, shifted vertically as described in the text. The
dash-dotted lines represent the best-fit CMFs over the relevant mass
range. The panel indicators refer to Fig. \ref{smcamd2.fig}.}
\end{figure*}

In order to achieve this goal, we present the CMFs for subsets of our
SMC cluster sample in Fig.  \ref{smcclfs4.fig}, where the cluster
subsamples were selected based on their age distributions. The age
limits used to generate the different panels in
Fig. \ref{smcclfs4.fig} are shown as the vertical dash-dotted lines in
Fig. \ref{smcamd2.fig}; the various subsets are also cross-linked
between the figures.

A closer look at Fig. \ref{smcamd2.fig} reveals that, because of the
variation in the observational detection limit as a function of age,
the lower-mass limits of our cluster subsamples differ. Thus, the CMFs
presented in Fig. \ref{smcclfs4.fig} are statistically complete above
different mass limits, as indicated by the horizontal dash-dotted
lines in Fig. \ref{smcamd2.fig} and the vertical dotted lines in
Fig. \ref{smcclfs4.fig}.

In all panels of Fig. \ref{smcclfs4.fig}, we have overplotted CMFs
with the canonical slope of $\alpha = 2$ (corresponding to a slope of
$-1$ in units of ${\rm d} \log(M_{\rm cl}/{\rm M}_\odot) / {\rm d}
\log (N_{\rm cl})$, used in these panels). We have only shifted and
scaled these lines vertically, as justified below. 

{\it We emphasise that for the star cluster infant mortality study
performed here, we need to choose the age ranges of our cluster
subsamples carefully}, for both physical reasons and also because of
the discrete nature of the model isochrones. Regarding the latter, it
is well known that broad-band SED fitting results in artefacts in the
cluster age distribution. This is predominantly caused by specific
features in the SSP models, such as the onset and presence of red
giant branch or asymptotic giant branch (AGB) stars at, respectively,
$\sim 10$ and $\sim 100$ Myr (e.g., Bastian et al. 2005).
Alternatively, both the age-metallicity and the age-extinction
degeneracies will affect the resulting cluster age distributions, thus
also leading to artefacts in the data (e.g., de Grijs et al. 2003b;
Anders et al. 2004). We have attempted to avoid placing our age range
boundaries around ages (and, where possible, have taken account of the
uncertainties in age in doing so) where the effects of such artefacts
might seriously impede the interpretation of the results. For
instance, one can see a clear artefact in the cluster age distribution
(which we will refer to as a ``chimney'') at $\log( {\rm Age / yr} )
\simeq 7.2 (\simeq 16$ Myr); the average uncertainties for these ages
are of order a few Myr, so that we decided to limit our youngest
cluster subsample to clusters younger than 10 Myr. If, instead, we had
adopted an age limit at $\log( {\rm Age / yr} ) = 7.17 (15$ Myr), we
would have had marginally better statistics, but our analysis would be
affected by the unknown effects of the age uncertainties associated
with this chimney (see Goodwin et al., in prep., for a detailed
discussion of the issues involved).

The rationale for adopting as our youngest subsample all clusters with
ages $\le 10$ Myr is that at these young ages, the vast majority of
the star clusters present will still be detectable, even in the
presence of early gas expulsion (e.g., Goodwin \& Bastian 2006) -- as
long as they are optically conspicuous. The CMF of this subsample is
shown in Fig. \ref{smcclfs4.fig}a.

Fig. \ref{smcclfs4.fig}c includes our sample clusters with ages in
excess of 40 Myr, up to 160 Myr. While the upper age limit ensures the
full inclusion of the clusters affected by the onset of the AGB stage,
its exact value is rather unimportant for our analysis, and it was
mainly determined by the need to have reasonable statistics in this
(and the upper) age range in Fig. \ref{smcclfs4.fig}d. The lower age
limit of this subsample is crucial, however. As shown by Goodwin \&
Bastian (2006), most dissolving clusters will have dispersed by an age
of $\sim 30$ Myr, while the surviving clusters will have returned to
an equilibrium state by $\sim 40$ Myr, when some of the early
expansion will have been reversed, depending on the effective
star-formation efficiency. This latter age is therefore a good lower
boundary to assess the surviving star cluster population.

We explicitly exclude any star clusters aged between 10 and 40 Myr
from our analysis. In this age range, which is shown in
Fig. \ref{smcclfs4.fig}b for completeness, it is likely that
dissolving star clusters that will not survive beyond about 30--40 Myr
might still be detectable and therefore possibly contaminate our
sample. In addition, this is the age range in which early gas
expulsion causes rapid cluster expansion, before settling back into
equilibrium at smaller radii; because of the expanded nature of at
least part of the cluster sample, we might not be able to detect some
of the lower-luminosity (and hence lower-mass) clusters that would
again show up beyond an age of $\sim 40$ Myr. At the same time, the
effects of ``infant weightloss'' (Weidner et al. 2007) will further
confuse the analysis in this age range (see Section \ref{infant.sec}
for details).

\section{Is there evidence for cluster infant mortality in the SMC?}
\label{discussion.sec}

\subsection{Young and intermediate-age clusters}

In Fig. \ref{smcclfs4.fig}a, we have included the best-fitting CMF
slope (dash-dotted line), in addition to the canonical $\alpha = 2$
CMF slope (dashed line). Both slopes are the same, within the
uncertainties. This also shows that the SMC's CMF at the youngest ages
is consistent with an $\alpha = 2$ power law down to cluster masses of
$\sim 125$ M$_\odot$, within the (Poissonian) uncertainties.

In the simplest case, in which the cluster formation rate has remained
roughly constant throughout the SMC's evolution (see, e.g., Boutloukos
\& Lamers 2003, their fig. 10; see also Gieles et al. 2007), the
number of clusters would simply scale with the age range covered. In
Figs. \ref{smcclfs4.fig}b, c and d, we show the canonical $\alpha=2$
CMF scaled from the best-fitting locus in Fig. \ref{smcclfs4.fig}a by
the difference in (linear) age range between the panels. The main
uncertainties introduced by this method are (i) fluctuations caused by
small-number statistics in the youngest age range (the effects of
which will be enhanced when scaling the young-age CMF to a greater age
range), and (ii) the exact length of the youngest age range
(especially considering the necessarily short extent of our youngest
age bin). While our {\sc galev} SSP models start at an age of 4 Myr,
the actual ages of a small subset of our sample clusters might be
somewhat younger. This introduces an artificial concentration of
clusters at our youngest model age, as can be seen in
Fig. \ref{smcamd2.fig}. It is, unfortunately, not straightforward to
remedy this situation based on broad-band imaging observations
alone. However, we note that it may take up to 3--5 Myr for an
embedded cluster to clear a cavity in its natal gas cloud for its
stars to become visible at optical wavelengths (see, e.g., Plante \&
Sauvage 2002 for a review of embedded young massive star cluster
observations). Therefore, we adopt the conservative working assumption
that our youngest age range runs from 3--10 Myr. The scaling from the
youngest age bin to that covering [40,160] Myr
(Fig. \ref{smcclfs4.fig}c, i.e., the most important age range for our
CMF comparison in the context of our infant mortality analysis) is
therefore a factor of $\sim 17$, or $\Delta \log (N_{\rm cl}) =
1.234$.

Despite the caveat regarding the absence of embedded star clusters in
our youngest subsample, we argue that this has a negligible effect on
the CMF presented in Fig. \ref{smcclfs4.fig}a, because of their very
small number. Recent, homogeneous observations of the SMC using the
{\sl Spitzer Space Telescope} in a number of mid-infrared passbands
(Bolatto et al. 2007) have shown that the vast majority of the
embedded sources are low-mass ($\ll 100$ M$_\odot$) young stellar
objects rather than more massive clusters and associations (we point
out that for the comparison done here, we are mostly interested in
clusters with masses greater than $10^3$ M$_\odot$). The possible
exceptions to this rule are few, and include the four youngest $\sim
3$ Myr-old SMC clusters, NGC 299, NGC 346, NGC 376, and NGC 602 (e.g.,
Sabbi et al. 2007).

The scaled canonical CMF in Fig. \ref{smcclfs4.fig}c is an almost
perfect fit to the observational CMF. Although the best-fitting CMF
slope is ${\rm d} \log(M_{\rm cl}/{\rm M}_\odot) / {\rm d} \log
(N_{\rm cl}) = -0.82 \pm 0.18$, this compares to ${\rm d} \log(M_{\rm
cl}/{\rm M}_\odot) / {\rm d} \log (N_{\rm cl}) = -1.01 \pm 0.20$ if we
ignore the lowest-mass clusters at $\log(M_{\rm cl}/{\rm M}_\odot) \le
3.2$, where there may be residual incompleteness effects (see the
selection area in Fig. \ref{smcamd2.fig} compared to the age-dependent
detection limit).

This very good match between the observed CMF for the age range from
40--160 Myr (Fig. \ref{smcclfs4.fig}c) and the scaled CMF from
Fig. \ref{smcclfs4.fig}a implies that {\it the SMC cluster system has
not been affected by any significant amount of cluster infant
mortality for cluster masses greater than a few} $\times 10^3$
M$_\odot$. Based on a detailed assessment of the uncertainties in both
the CMFs and the age range covered by our youngest subsample, we can
limit the extent of infant mortality between the youngest and the
intermediate age range to a maximum of $\lesssim 30$ per cent
($1\sigma$). We rule out a $\sim 90$ per cent (infant) mortality rate
per decade of age at a $>6 \sigma$ level. This result is in excellent
agreement with that of Gieles et al. (2007); it is, however, in direct
contradiction to the claim of Chandar et al. (2006) that the SMC
cluster system has undergone sustained destruction at very high rates
(up to 90 per cent per decade in logarithmic age) for the full age
range up to $\sim 1$ Gyr, although we note that Chandar et al. (2006)
do not include the youngest SMC clusters in their analysis.

As an important caveat, we remind the reader that the main underlying
assumption leading to this result is the notion that the SMC's CFR has
been approximately constant over the time-scale of $\sim 1$ Gyr. If
this were seriously in error, in order for this to give rise to the
result reported here, the SMC's average CFR must have been
significantly enhanced in the $40 - 160$ Myr-old age range, by up to
an order of magnitude, compared to that at present. There is no clear
evidence, in either the current data set or the earlier work by
Boutloukos \& Lamers (2003; see also Gieles et al. 2007), nor in the
age distribution of the field stars (Chandar et al. 2006; Chiosi \&
Vallenari 2007), to suggest that this is the case. In fact, if
anything, we might expect an enhanced CFR around the time of the last
close encounter between the SMC and the LMC, some $200 - 500$ Myr ago
(see, e.g., Heller \& Rohlfs 1994; Gardiner \& Noguchi 1996; see also
Chiosi et al. 2006, but see Chiosi \& Vallenari 2007 for an
alternative interpretation), i.e., significantly longer ago than the
age range probed by our intermediate-age clusters in
Fig. \ref{smcclfs4.fig}b.

For completeness, we also show the best-fit power-law CMF, as well as
the scaled canonical CMF, in Fig. \ref{smcclfs4.fig}b for the age
range between 10 and 40 Myr. Although {\it we strongly caution against
placing too much significance on the analysis of the CMF in this age
range}, for the reasons outlined in Section \ref{analysis.sec}, it is
interesting to note that the scaled canonical CMF does in fact seem to
describe the extremities of this CMF reasonably well. However, at
intermediate masses (a few $\times 10^3 - 10^4$ M$_\odot$) the
observed CMF exceeds the theoretical prediction for a constant cluster
formation rate. Although the effects of cluster expansion and infant
weightloss most likely contribute to confusing the emerging picture,
the main cause of this discrepancy is owing to the artificial chimney
at $\log( {\rm Age / yr} ) \simeq 7.2$. The cluster numbers in this
age range are dominated by this artefact, so it is important that we
understand in which sense this affects our results. Because of the
discreteness of the isochrones in our SSP models around this age, and
the tendency for the broad-band fitting routine to iterate to a local
minimum $\chi^2$ solution, most (but not all) of the clusters in this
chimney should have been assigned somewhat greater ages. Because their
ages have been underestimated (by up to 0.1--0.2 dex in logarithmic
age), the associated masses have also been underestimated, by a
similar amount. Unfortunately, until more detailed SSP models become
available, there is no easy way out. However, a qualitative assessment
suggests that if we were able to correct for this chimney, the derived
masses of at least a fraction of the clusters affected would be
greater, and thus that the apparent excess in Fig. \ref{smcclfs4.fig}b
would be redistributed towards greater masses. The result would be a
smoother CMF, more akin to the scaled canonical $\alpha=2$ CMF of
Fig. \ref{smcclfs4.fig}a. 

The alternative interpretation, i.e., that the star cluster formation
rate in the SMC has undergone a significant increase in the age range
between 10 and 40 Myr appears to be effectively ruled out by the
complementary analysis of Chiosi et al. (2006). In fact, these authors
find evidence suggesting the contrary, i.e., that the SMC cluster
population has seen periods of enhancement during the first $\sim 15$
Myr, and at around 90 Myr -- this implies that in the age range of
interest here, the cluster formation rate found by Chiosi et
al. (2006) was in fact {\it reduced} with respect to the most recent
cluster forming episode (roughly equivalent to the cluster sample
shown in Fig. \ref{smcclfs4.fig}a). Equivalently, neither Rafelski \&
Zaritsky (2005), nor either Chandar et al. (2006) or Gieles et
al. (2007) find an episode of increased cluster formation at a few
$\times 10^7$ yr, despite the fundamentally different conclusions
drawn by the authors of the latter two studies.

\subsection{The oldest sample clusters}

Finally, in Fig. \ref{smcclfs4.fig}d we show the combined SMC CMF for
clusters from 160 Myr up to 1.0 Gyr, as well as the scaled canonical
CMF. The latter matches the highest-mass ($\log(M_{\rm cl}/{\rm
M}_\odot) \ga 4.8$) part of the observed CMF, but significantly
overpredicts the number of lower-mass ($\log(M_{\rm cl}/{\rm M}_\odot)
\lesssim 4.8$) clusters. This flattening of the CMF with respect to
our intermediate age range (Fig. \ref{smcclfs4.fig}c) evidences the
increased importance of mass-dependent cluster disruption, as
described in detail by, e.g., Lamers et al. (2005; see specifically
their fig. 4). However, we note that the calculations of Lamers et
al. (2005) are based on coeval stellar populations, while our oldest
age bin contains a mixture of differently aged clusters. Nevertheless,
it is apparent from fig. 4 of Lamers et al. (2005) that the flattening
of the CMF increases as a function of age, starting at early ages. In
a mixed intermediate-age population, the individual roughly coeval
subpopulations giving rise to the overall CMF therefore lead to a
flattened CMF with respect to the initial cluster mass distribution
(which we have shown in the SMC to be equivalent to the canonical
$\alpha=2$ power-law CMF; see Fig. \ref{smcclfs4.fig}a). A comparison
of our observations with the Lamers et al. (2005) calculations is
therefore relevant to first order. A similar flattening of the CMF
with increasing age is also predicted as owing to the effects of the
underlying galactic tidal field (e.g., Vesperini \& Heggie 1997;
Baumgardt \& Makino 2003). We will discuss these older clusters in
more detail in a follow-up paper (Goodwin et al., in prep.), in which
we will discuss the entire evolutionary sequence of the star cluster
systems of both Magellanic Clouds, and where we aim to understand the
{\it physics} driving the early evolution of these star clusters
systems.

\subsection{Additional supporting evidence for little early disruption}

\begin{figure}
\psfig{figure=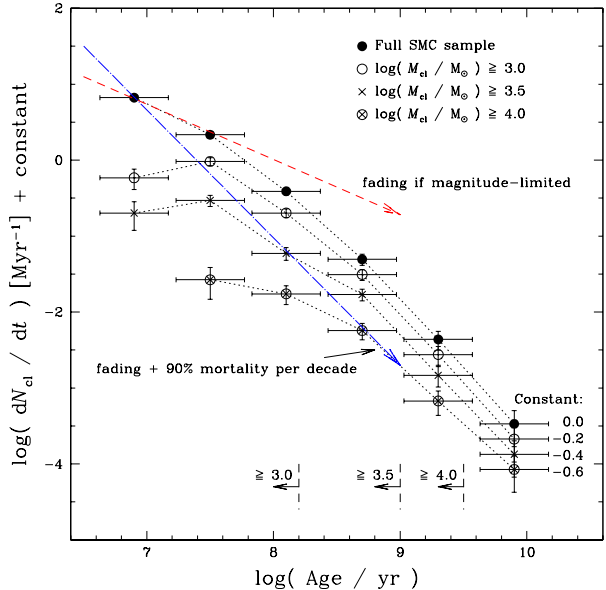,width=\columnwidth}
\caption{\label{dndt.fig}Age distribution of the SMC cluster sample in
units of cluster number per Myr. We show four different (sub)samples,
including the full magnitude-limited SMC sample, and three
mass-limited subsamples, as indicated in the figure legend. The
mass-limited subsamples have been shifted vertically by the constant
offsets indicated on the right-hand side for reasons of clarity
(without these offsets, the data points would all overlap at the
oldest ages). For that same reason, we have also connected the data
points for each of the (sub)samples. The mass-limited subsamples are
$\ge 50$ per cent complete to the left of the vertical dashed lines at
the bottom of the figure, where the numbers refer to the 50 per cent
completeness limits for a given range, expressed in $\log( M_{\rm cl}
/ {\rm M}_\odot )$. The vertical error bars are simple Poissonian
errors; the horizontal error bars indicated the age range used for the
generation of these data points. Finally, the dashed arrow shows the
expected effects due to fading of a cluster sample made up of SSPs,
based on the {\sc galev} SSP models (see also Gieles et al. 2007); the
dash-dotted arrow represents the combined effects of a fading cluster
population and 90 per cent cluster disruption per decade in $\log(
\mbox{Age yr}^{-1} )$, up to ages of 1 Gyr.}
\end{figure}

In Fig. \ref{dndt.fig} we show the SMC cluster age distribution
expressed in units of ${\rm d}N_{\rm cl} / {\rm d}t$, i.e., the number
of cluster per unit time-scale (for which we adopt $10^6$ yr). To
first order, our age distribution is similar to that based on the
Rafelski \& Zaritsky (2005) data used by both Chandar et al. (2006)
and Gieles et al. (2007). It is also roughly similar to the age
distribution derived independently by Chiosi et al. (2006).

Gieles et al. (2007) analysed their cluster age distribution very
carefully, and found little evidence for infant mortality in the SMC
cluster system. They showed that the decline in ${\rm d}N_{\rm cl} /
{\rm d}t$ in their sample could be attributed entirely to evolutionary
fading of the cluster population, irrespective of which SSP models are
used. They derived that for a magnitude-limited sample, as also
discussed in this paper, the decline in ${\rm d}N_{\rm cl} / {\rm d}t$
as a function of age is graphically described by a slope of $-0.72$,
if the cluster ages are based on the {\sc galev} SSP models. In
Fig. \ref{dndt.fig}, we show the expected effects of evolutionary
fading of the cluster population as the dashed arrow. It is
immediately clear that the decline in the age distribution up to
$\log( \mbox{Age yr}^{-1} ) \simeq 7.8$ can indeed be entirely
attributed to evolutionary fading. This is supported by the
mass-limited subsamples shown in Fig. \ref{dndt.fig}: for all mass
ranges, they show an essentially flat age distribution up to $\sim
10^8$ yr. This is in support of the results of both Gieles et
al. (2007), and Chandar et al. (2006; their fig. 1), although the
latter authors favoured a different interpretation.

The age distribution at $\log( \mbox{Age yr}^{-1} ) \simeq 8.2 \pm
0.3$ (covering the age range from about 80 to 320 Myr) falls below the
fading line, however. For a constant cluster formation rate over this
entire period, and if we normalise the age distribution at our
youngest data point, we would need the SMC cluster sample to have
suffered from $\sim 20-50$ per cent disruption in order to match the
observations. We can firmly rule out a constant $\sim 90$ per cent
disruption rate per decade in age, up to an age of 1 Gyr. The expected
effects of evolutionary fading combined with a 90 per cent disruption
rate are shown as the dash-dotted arrow in Fig. \ref{dndt.fig}. The
arrow clearly does not fit the observed age distribution, if we
require it to pass through our youngest data point. We note, however,
that the slope of this latter arrow is very similar to that of the age
distribution of the full SMC sample for ages in excess of a few
$\times 10^8$ yr, when secular disruption is likely to take over. We
will discuss these older age ranges in detail in Goodwin et al. (in
prep.).

In summary, we set out to shed light on the controversy surrounding
the early evolution and disruption of star clusters in the SMC. We
embarked on a fresh approach to the problem, using an independent,
homogeneous data set of $UBVR$ imaging observations, from which we
obtained the cluster age distribution in a self-consistent manner. We
conclude that the optically selected SMC star cluster population has
undergone at most $\sim 30$ per cent ($1\sigma$) ``infant
mortality''. Using the age distribution of the SMC cluster sample in
units of the number of clusters observed per unit time-scale, we
independently confirm this scenario. Gieles et al. (2007) reached a
similar conclusion.

\section*{acknowledgments} We thank Peter Anders for performing the
SED cluster fits using his {\sc AnalySED} tool. We are also grateful
to the anonymous referee for a careful reading of this paper and for
suggesting useful additional lines of investigation. RdG acknowledges
the discussions at the ESO workshop on ``12 Questions in Star and
Massive Star Cluster Formation'' organised by Markus Kissler-Patig and
Tom Wilson. RdG and SPG also acknowledge partial financial support
from the Royal Society in the form of a UK-China International Joint
Project, as the revised version of this paper was prepared during a
collaborative visit to China. We are grateful to an anonymous
independent referee for suggesting a number of improvements to the
paper that strengthened the main results. This research has made use
of NASA's Astrophysics Data System Abstract Service.

\end{document}